\documentclass[sigconf]{acmart}

\AtBeginDocument{%
  }

\usepackage{multicol}
\usepackage{multirow}
\usepackage{tabularx}
\usepackage{float}
\usepackage{makecell}
\usepackage{booktabs}
\usepackage{siunitx}
\usepackage{array}
\usepackage{amsmath}
\usepackage{pgfplots}
\pgfplotsset{compat=1.18}

\usepackage{bm}

\usepackage{fancyhdr}
\renewcommand\footnotetextcopyrightpermission[1]{} 

\begin{document}

\title{Cross-Domain Sequential Recommendation via Neural Process}

\author{Haipeng Li}
\authornote{Equal contributions to this work.}
\affiliation{%
  \institution{Nanjing University of Science and
Technology}
  \city{Nanjing}
  \country{China}
}
\email{lihaipeng@njust.edu.cn}

\author{Jiangxia Cao}
\authornotemark[1]
\affiliation{
    \institution{Kuaishou Technology}
      \city{Beijing}
  \country{China}
}
\email{caojiangxia@kuaishou.com}

\author{Yiwen Gao}
\affiliation{
\institution{Nanjing University of Science and
Technology}
  \city{Nanjing}
  \country{China}
}
\email{gaoyiwen@njust.edu.cn}

\author{Yunhuai Liu}
\affiliation{%
 \institution{Peking University}
       \city{Beijing}
  \country{China}
}
\email{Yunhuai.liu@pku.edu.cn}

\author{Shuchao Pang}
\authornote{Corresponding authors.}
\affiliation{
\institution{Nanjing University of Science and
Technology}
      \city{Nanjing}
  \country{China}
}
\email{pangshuchao@njust.edu.cn}

\begin{abstract}
    Cross-Domain Sequential Recommendation (CDSR) is a hot topic in sequence-based user interest modeling, which aims at utilizing a single model to predict the next items for different domains.
    To tackle the CDSR, many methods are focused on domain overlapped users' behaviors fitting, which heavily relies on the same user's different-domain item sequences collaborating signals to capture the synergy of cross-domain item-item correlation.
    Indeed, these overlapped users occupy a small fraction of the entire user set only, which introduces a strong assumption that the small group of domain overlapped users is enough to represent all domain user behavior characteristics.
    However, intuitively, such a suggestion is biased, and the insufficient learning paradigm in non-overlapped users will inevitably limit model performance.
    Further, it is not trivial to model non-overlapped user behaviors in CDSR because there are no other domain behaviors to collaborate with, which causes the observed single-domain users' behavior sequences to be hard to contribute to cross-domain knowledge mining.
    Considering such a phenomenon, we raise a challenging and unexplored question: How to unleash the potential of non-overlapped users' behaviors to empower CDSR?   
    To this end, we propose a novel CDSR framework with neural processes (NP), briefly termed \textbf{CDSRNP}, where NP combines the advantages of meta-learning and stochastic processes.
    As a meta-learning based method, we first sample some observed overlapped users' behaviors as the support set to empower query users' prediction.
    Next, we employ the NP principle to align the cross-domain correlation prior/posterior distributions generated by support/query user sets, thus the query user (e.g., non-overlapped user) behaviors sequence could also establish a straight bridge to connect other domain items.
    Additionally, we design a fine-grained interest adaptive layer to identify the users' interests to enhance prediction.
    Experimental results illustrate that CDSRNP outperforms state-of-the-art methods in two real-world datasets.
\end{abstract}




\settopmatter{printfolios=true}

\maketitle

\section{Introduction}
\begin{sloppypar} 
    Sequential Recommendation (SR)\cite{bert4rec,hidasi2015session,xie2022contrastive} plays an increasingly important role in various digital platforms to capture user's dynamic interest change, especially in e-commerce shopping and entertainment short-video platforms.
    In the real world, when users experience that platform, some high-active users will interact with multiple domain items, e.g., buying books/movies at Amazon or watching short videos/live-streaming on TikTok.
    Therefore, a domain-specific SR model is hard to learn through user preferences but easily captures biased user preferences in a single domain.
    To this end, the Cross-Domain Sequence Recommendation (CDSR)\cite{wang2023linsatnet,wang2020combinatorial,scarselli2008graph,xu2024towards,lin2024mixed} was proposed, which extends SR to a cross-domain setting to consider temporal dynamics of the user's domain-shared interests transferring.
    Moreover, the CDSR paradigm enables robust performance even in the presence of limited user interaction (e.g., data-rich domain bootstraps data-scarce domain performance), thus effectively mitigating the problem of data sparsity in recommendation systems.

    To our knowledge, in the training procedure, previous CDSR methods\cite{lin2024mixed,li2021dual,xu2024towards} are focused on domain--overlapped users' behavior sequence, which heavily relies on the same user's different-domain item sequences collaborating signals to capture the synergy of cross-domain item-item correlation.
\begin{figure}[ht]
  \centering
  \includegraphics[width=\linewidth]{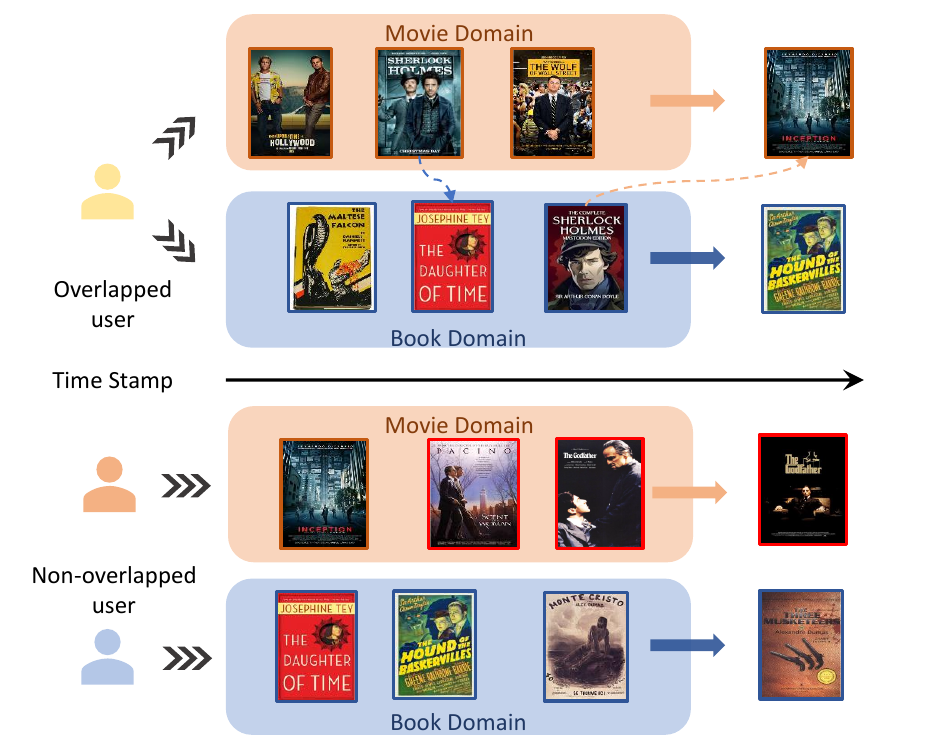}
  \caption{An illustration of overlapped/non-overlapped users, where previous methods focus on the overlapped users' behavior learning.}
  \label{fig:liuchnegtu1}
  \Description{}
\end{figure}
    For instance, the C$^2$DSR\cite{cao2022contrastive} only utilizes the overlapped users' single-domain and cross-domain sequences to generate domain-specific and domain-shared representations with an overlapped users-based in-batch contrastive module as show in Fig. \ref{fig:liuchnegtu1}.
    Although effective, however, such paradigm approaches are based on the assumption that the user-item interaction pattern of overlapped and non-overlapped users are the same: the small group
    domain-overlapped users are enough to represent all domains
    user behaviors characteristics.
    Actually, these overlapped users occupy a small fraction of the entire user set only, thus the insufficient learning paradigm in non-overlapped users will inevitably lead to models’ performance degradation.
    Nevertheless, it is not trivial to model non-overlapped user behaviors in CDSR because there are no other domain behaviors to collaborate with, which causes the observed single-domain behavior sequences to be challenged to contribute to cross-domain knowledge mining.
    
    Intuitively, the final goal of the CDSR model is to replace all other single-domain SR models. Thus, the non-overlapped user's behavior sequence should be considered, not only to learn comprehensive and worthwhile single-domain item-item collaborate signal but also to capture the complete all domains user-item interaction pattern distribution.
    Considering such a phenomenon, we raise a challenging and unexplored question: \textbf{How to unleash the potential of non-overlapped user’s behaviors to empower CDSR?} 
    In this paper, we propose a novel CDSR framework with neural processes, briefly termed \textbf{CDSRNP}. 
    Neural processes represent an innovative approach to meta-learning \cite{bharadhwaj2019meta,Mamo,Melu,lu2020meta} that combines the strengths of neural networks and stochastic processes to establish a powerful function.
    To be specific, CDSRNP follows a meta-learning workflow, and we first sample some observed overlapped users' behaviors as a support set to empower query overlapped or non-overlapped user prediction.
    Next, we employ the NP principle to learn the prior sample distribution in the support set as well as the posterior sample distribution in the query set. 
    By aligning those prior and posterior distributions, the query user (e.g., non-overlapped user) behaviors sequence item could establish a straight bridge to other domains item sequence item.
    Additionally, instead of learning the domain-specific and domain-shared characteristics, we also design a fine-grained interest adaptive layer to identify the personalized highly related support users' interests to transfer beneficial information for enhancing query user prediction.
    Extensive experiments show that CDSRNP achieves significant improvements over state-of-the-art baselines on two real-world datasets.
    In summary, our contributions are as follows:
    \begin{itemize}  
        \item We introduce a fresh perspective to enhance CDSR by further considering the non-overlapped user behaviors, which sheds light on building a new paradigm for CDSR.
        \item We devise a novel approach, CDSRNP, which could capture the complete overlapped and non-overlapped user interaction pattern distribution. Further, we introduce a fine-grained interest adaptive layer to detect related overlapped users' interests for non-overlapped user prediction.
        \item We perform extensive experiments and detailed analysis on two real-world datasets to evaluate model performance, which demonstrates that our CDSRNP achieves consistent and significant improvements over state-of-the-art baselines.
    \end{itemize}
\end{sloppypar}

\section{PRELIMINARY}

\begin{figure}[t]
  \centering
  \includegraphics[width=\linewidth]{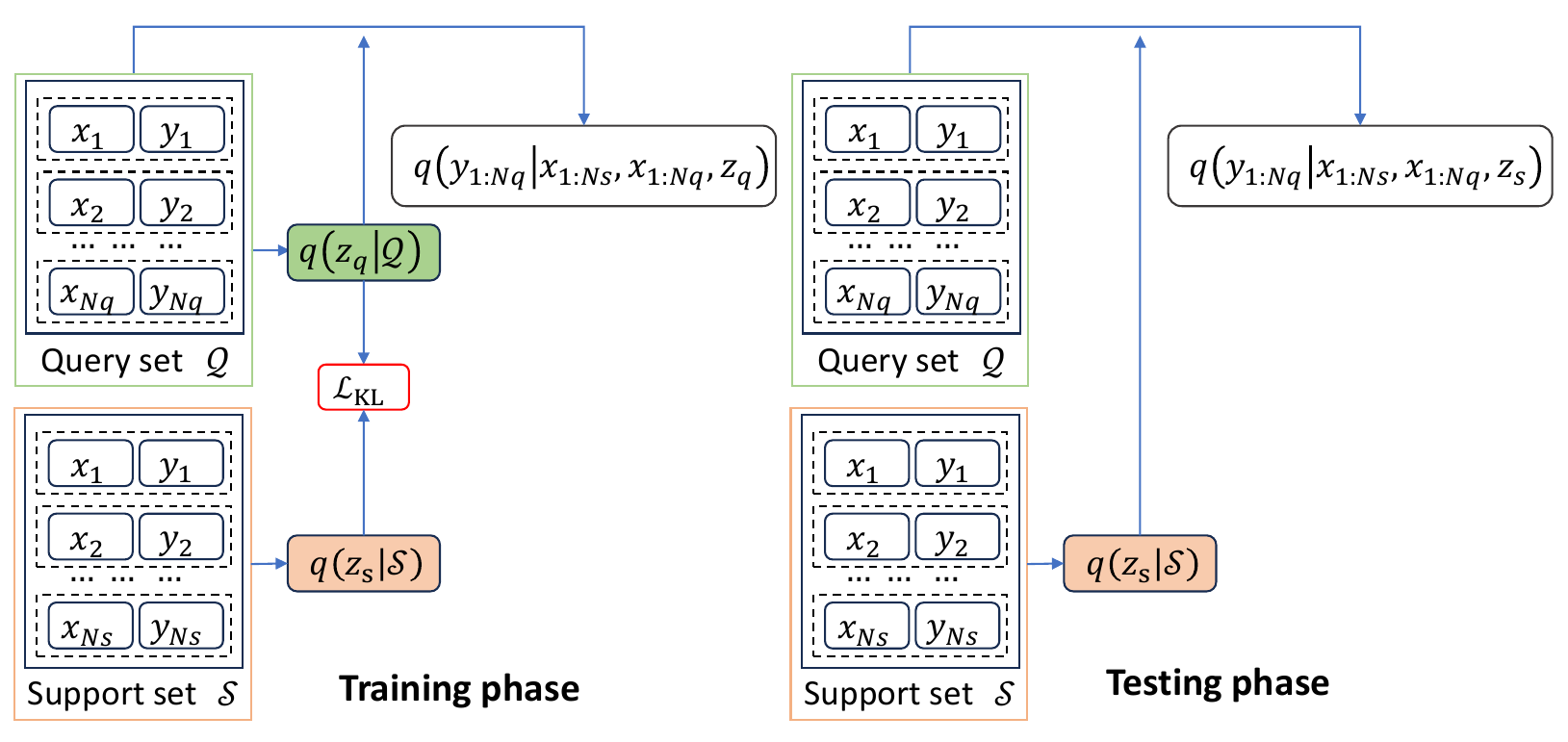}
  \caption{Neural processes in the training/testing phases}
  \label{fig:liuchnegtu2}
  \Description{}
\end{figure}

\subsection{Neural Processes}
    Neural Processes (NP) \cite{garnelo2018conditional,garnelo2018neural} learns an approximation of a stochastic process based on neural networks.
    Neural networks use observations as sample points for a stochastic process to obtain a posterior distribution. 
    This posterior distribution approximates the implied prior distribution.
    By utilizing this posterior distribution, neural networks can construct a function that transforms inputs from the input space into corresponding outputs in the output space.
    This approach enables neural networks to be applied to various tasks, including generative tasks\cite{garnelo2018neural,garnelo2018conditional,kim2019attentive}, image recognition\cite{wang2022np}, and cross-domain recommendations\cite{lin2021task}. 
    We briefly summarize the information as the input data
    $\mathbf{x}_{1:n}$, the output label ${y}_{1:n}$. 
    Given a particular instantiation of the stochastic process $\mathbf{f}$, 
    the joint distribution is defined as:
    \begin{equation}
        \rho_{\mathbf{x}_{1:n}}({y}_{1:n})=\int p(\mathbf{f}_i)p({y}_{1:n}|\mathbf{f},\mathbf{x}_{1:n})d\mathbf{f}_i,
        \label{eq:1}
    \end{equation}
    where $\rho$ denotes the abstract probability distribution of all random variables.
    To be able to represent the stochastic process using NP, we use a neural network approximation to represent the stochastic process $\mathbf{F}(\mathbf{x}) = h(\mathbf{x},\mathbf{z})$, parameterized with a high-dimensional vector $\mathbf{z}$.
    We obtain the high-dimensional hidden variable $\mathbf{z}$ from the standard Gaussian distribution and add noise to it (introducing randomness).
    Since each stochastic process is independent of each other, we can rewrite Eq.\eqref{eq:1}:
    \begin{equation}
        \rho(\mathbf{z},{y}_{1:n})|\mathbf{x}_{1:n})=  p(\mathbf{z})\prod_{i=1}^{n}p({y}_i|h(\mathbf{x}_i,\mathbf{z}),\mathbf{\sigma}^{2}).
        \label{eq:2}
    \end{equation}
    Next, learning to generate a neural network encoder, parameterizing the variable $\mathbf{z}$.
    Let $q(\mathbf{z}|\mathbf{x}_{1:n},{y}_{1:n})$ be a variational posterior of the latent variables $\mathbf{z}$.
    Using the evidence lower-bound(ELBO) the following equation can be obtained:
    \begin{equation}
        % \small
        \begin{aligned}
        &\log p({y}_{1:n}|\mathbf{x}_{1:n}) \\
        &\geqslant E_{q(\mathbf{z}|\mathbf{x}_{1:n},{y}_{1:n})}\left[\sum_{i=1}^{n}\log p({y}_{i}|\mathbf{x}_{i},\mathbf{z})+\log\frac{p(\mathbf{z})}{q(\mathbf{z}|\mathbf{x}_{1:n},{y}_{1:n})}\right].
        \end{aligned}
        \label{eq:3}
    \end{equation}
    For the model to make better predictions, we divide the dataset into support set 
    $\mathcal{S}(\mathbf{x}_{1:m},{y}_{1:m}$) and query set $\mathcal{Q}(\mathbf{x}_{m+1:n},{y}_{m+1:n}$).
    Let the model predict the content of the query set given the support set and 
    we rewrite the above equation:
    \begin{equation}
        \begin{aligned}
        &\log p({y}_{q}|\mathbf{x}_{1:n},{y}_{1,m}) \\
        &\geqslant E_{q(\mathbf{z}|\mathbf{x}_{1:n},{y}_{1:n})}\left[\sum_{i=m+1}^{n}\log p({y}_{i}|\mathbf{x}_{i},\mathbf{z})+\log\frac{p(\mathbf{z}|\mathbf{x}_s,{y}_s)}{q(\mathbf{z}|\mathbf{x}_{1:n},{y}_{1:n})}\right].
        \end{aligned}
        \label{eq:4}
    \end{equation}
    Since the conditional prior $p(\mathbf{z}|\mathbf{x}_{s},{y}_{s})$ is difficult to obtain, we use the conditional posterior $q(\mathbf{z}|\mathbf{x}_{s},{y}_{s})$ approximation instead.
    We can rewrite the Eq.\eqref{eq:4} to get:
    \begin{equation}
        \begin{aligned}
        &\log p({y}_{q}|\mathbf{x}_{1:n},{y}_{1,m}) \\
        &\geqslant E_{q(\mathbf{z}|\mathbf{x}_{1:n},{y}_{1:n})}\left[\sum_{i=m+1}^{n}\log p(\mathbf{y}_{i}|\mathbf{x}_{i},\mathbf{z})+\log\frac{q(\mathbf{z}|\mathbf{x}_{s},{y}_{s})}{q(\mathbf{z}|\mathbf{x}_{1:n},{y}_{1:n})}\right]  \\
        &=\underbrace{E_{q(\mathbf{z}_q|\mathbf{x}_q,{y}_q)}\log p({y}_q|\mathbf{x}_q,\mathbf{z}_q))}_
        {log-likelihood}
        -\underbrace{KL(q(\mathbf{z}_q|\mathbf{x}_q,{y}_q)||q(\mathbf{z}_s||\mathbf{x}_s,{y}_s))}_{KL divergence}.
        \end{aligned}
        \label{eq:5}
    \end{equation}
    With the above formulas, we can observe two optimization objectives, the first one is the recommendation task of CDSR that is expected to be accomplished, and the second one is to compute the $KL$ triple to ensure that the support set and the query set are the same stochastic process.
    
    The training and testing process of the neural network is shown in Fig. \ref{fig:liuchnegtu2}.
    It is worth noting that $\mathbf{z}_q$ is used instead of $\mathbf{z}_s$ for prediction due to the lack of corresponding labels ${y}_q$ in the testing session. 
    This can be done because we ensure that the support set and the query set are the same stochastic process by $KL$ constraints during the training process.

\subsection{Problem Setting}
    This paper explores a partially overlapping CDSR scenario, aiming to provide recommendations for users based on observed user-item interactions across two domains.
    For illustration, we consider two domains ($A$ and $B$),
    let 
    $\mathcal{V}=(\mathcal{V^A},\mathcal{V^B})$
    represent the items set in two domains.
    $X^{A}=(v_1^{A},\ldots,v_T^{A})$ denotes user's interaction sequences in domain $A$. 
    For each user data sample $(X^{A}, X^{B}, v_{T+1}, y)$, we aim to predict the arbitrary item $v_{T+1}$ candidate is a positive or negative item.
   
\section{Methodology}
\begin{figure*}[ht]
\centering
\vspace{-0.5em} 
\includegraphics[width=\textwidth]{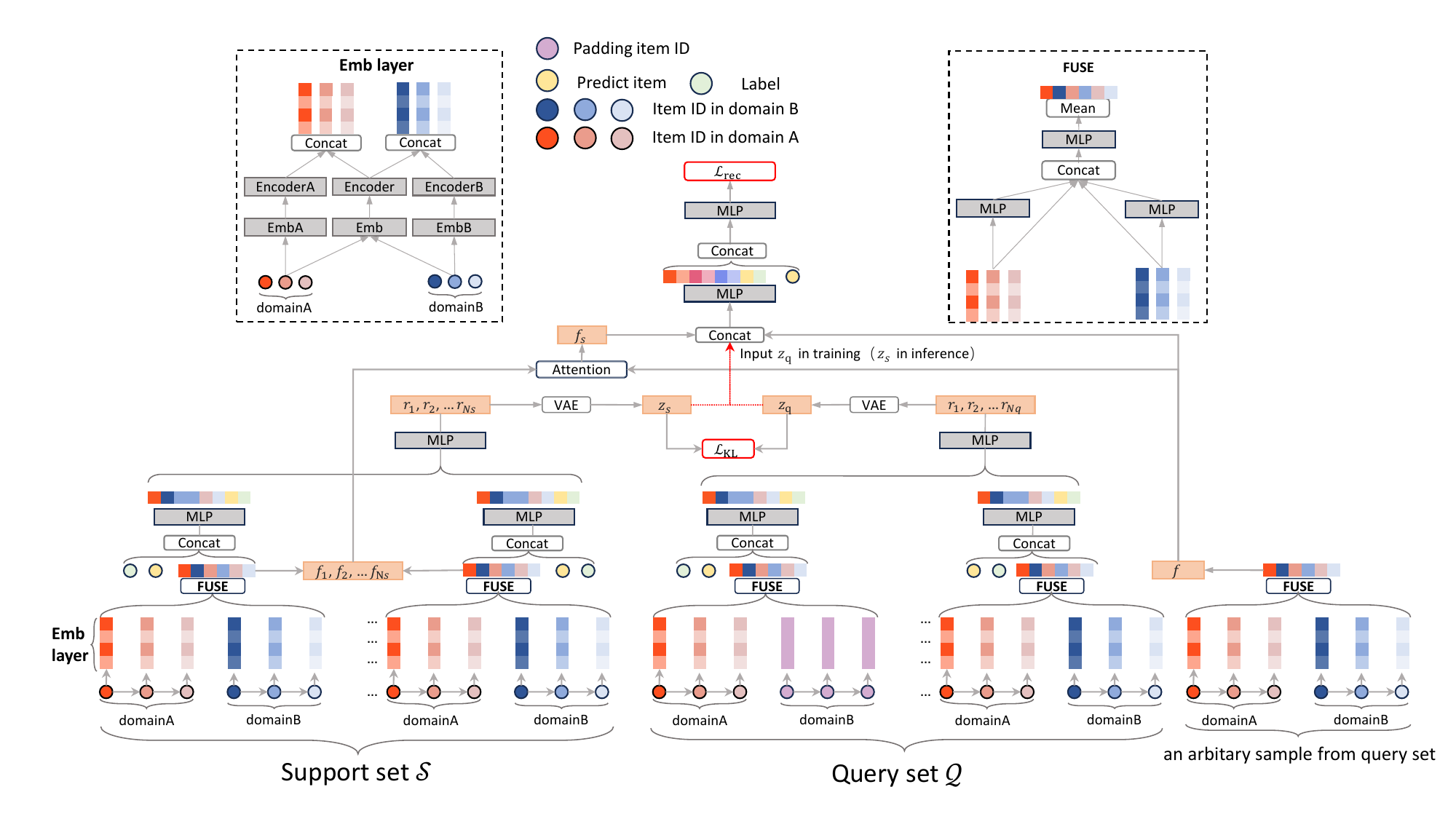}
\vspace{-0.8em}
\caption{The framework of CDSRNP in the training phase. $\mathbf{z}_s$ will be used instead of $\mathbf{z}_q$ in the testing phase}
\label{fig:moxingtu}
\Description{}
\end{figure*}

    This section introduces our model CDSRNP implementation of the neural process principle as shown in Fig. \ref{fig:moxingtu}.

\subsection{Embedding Layer}
    To effectively capture single-domain and cross-domain features, we use a specific and shared embedding module. 
    Specifically, we construct domain-specific local embeddings and cross-domain shared embeddings for items as:
    $\mathbf{V}^A \in\mathbb{R}^{|\mathcal{V}^A|\times D}$,
    $\mathbf{V}^B \in \mathbb{R}^{|\mathcal{V}^{B}|\times D}$ and
    $\bar{\mathbf{V}} \in \mathbb{R}^{|\mathcal{V}^{A}+\mathcal{V}^{B}|\times D}$,
    where $D$ denotes the latent dimension, to capture item features across two domains from a representation learning perspective.
    According to $\mathbf{V}^A, \mathbf{V}^{B}, \bar{\mathbf{V}}$, we could obtain the embeddings $(\mathbf{v}_1^A,\mathbf{v}_2^A,\ldots,\mathbf{v}_{T}^A)$/$(\bar{\mathbf{v}}_1^A,\bar{\mathbf{v}}_2^A,\ldots,\bar{\mathbf{v}}_{T}^A)$ for observed item sequence $(v_1^A,v_2^A,\ldots,v_{T}^A)$, and $(\mathbf{v}_1^B,\mathbf{v}_2^B,\ldots,\mathbf{v}_{T}^B)$/$(\bar{\mathbf{v}}_1^B,\bar{\mathbf{v}}_2^B,\ldots,\bar{\mathbf{v}}^B_{T})$ for observed item sequence $(v_1^B,v_2^B,\ldots,v_{T}^B)$~\footnote{It's important to note that sequences shorter than $T$ will be padded with a constant zero vector, e.g. $[\texttt{<pad>}, \texttt{<pad>}, \dots, \mathbf{v}_{T}^A]$ and  $[\texttt{<pad>}, \texttt{<pad>}, \dots, \mathbf{v}_{T}^B]$, where $\texttt{<pad>}$ denotes a $D$-dimension zero vector. For non-overlapping users, we pad the sequences of the other domain with an entire zero vector sequence.}.
    To further capture the position of items in the sequence, we integrate learnable positional embeddings into item embeddings, as follows:
\begin{equation}
\begin{aligned}
    \mathbf{E}^A_{\texttt{specific}} &= \left[\mathbf{v}_1^A,\mathbf{v}_2^A,\ldots,\mathbf{v}_{T}^A\right] + \left[\mathbf{p}_1^A, \mathbf{p}_2^A, \cdots, \mathbf{p}_{T}^A\right], \\
    \mathbf{E}^A_{\texttt{shared}} &= \left[\bar{\mathbf{v}}_1^A,\bar{\mathbf{v}}_2^A,\ldots,\bar{\mathbf{v}}_{T}^A\right] + \left[\mathbf{p}_1, \mathbf{p}_2, \cdots, \mathbf{p}_{T}\right], \\
    \mathbf{E}^B_{\texttt{specific}} &= \left[\mathbf{v}_1^B,\mathbf{v}_2^B,\ldots,\mathbf{v}_{T}^B\right] + \left[\mathbf{p}_1^B, \mathbf{p}_2^B, \cdots, \mathbf{p}_{T}^B\right], \\
    \mathbf{E}^B_{\texttt{shared}} &= \left[\bar{\mathbf{v}}_1^B,\bar{\mathbf{v}}_2^B,\ldots,\bar{\mathbf{v}}_{T}^B\right] + \left[\mathbf{p}_1, \mathbf{p}_2, \cdots, \mathbf{p}_{T}\right],
\end{aligned}
\label{eq:8}
\end{equation}
    where $\mathbf{E}^A_{\texttt{specific}},\mathbf{E}^A_{\texttt{shared}}
    \in\mathbb{R}^{T \times D}$ denote the specific and shared embeddings for domain $A$ , and $\mathbf{E}^A_{\texttt{specific}},\mathbf{E}^A_{\texttt{shared}}\in\mathbb{R}^{T \times D}$ for domain $B$. 
    Here $\mathbf{P}^A,\mathbf{P}^B$ and $\mathbf{P}$ are three
    learnable positional embeddings.

\subsection{Sequence Encoder}
    After obtaining the sequence input embedding, we next use a sequence encoder to learn sequence features in different domains.
    \begin{equation}
    \begin{split}
        \mathbf{X}_{\texttt{specific}}^A&=\texttt{SeqEncoder}_{\texttt{specific}}(\mathbf{E}^A_{\texttt{specific}}),\\
        \mathbf{X}_{\texttt{shared}}^{A}&=\texttt{SeqEncoder}_{\texttt{shared}}(\mathbf{E}_{\texttt{shared}}^A), \\
        \mathbf{X}_{\texttt{specific}}^B&=\texttt{SeqEncoder}_{\texttt{specific}}(\mathbf{E}^B_{\texttt{specific}}),\\
        \mathbf{X}_{\texttt{shared}}^{B}&=\texttt{SeqEncoder}_{\texttt{shared}}(\mathbf{E}_{\texttt{shared}}^B), \\
        \label{eq:10}
    \end{split}
    \end{equation}
    where the $\mathbf{X}_{\texttt{specific}}^A$, $\mathbf{X}_{\texttt{shared}}^{A}$, $\mathbf{X}_{\texttt{specific}}^B$, $\mathbf{X}_{\texttt{shared}}^{B} \in\mathbb{R}^{T\times D}$ are encoder results of item sequence representation to describe user interests, the $\texttt{SeqEncoder}(\cdot)$ is a sequence feature extractor, in CDSRNP, we implement it by a masked self-attention layer.
    
    By concatenating these extracted features, we obtain a comprehensive representation that encapsulates both the specific-domain and cross-domain user characteristics.
    To learn the distinctive features of users with various identities across different domains, we use specialized feature processing layers. 
    These layers aim to capture both the unique features within each domain and the nuanced differences between overlapping and non-overlapping user identities within the same domain.
    \begin{equation} 
    \begin{aligned}
    \mathbf{F}^{A} = \texttt{MLP}^A([&\mathbf{X}^A_{\texttt{specific}}\oplus\mathbf{X}^{A}_{\texttt{shared}}]), \\
    \mathbf{F}^{B} = \texttt{MLP}^B([&\mathbf{X}^B_{\texttt{specific}}\oplus\mathbf{X}^{B}_{\texttt{shared}}]), \\
    \mathbf{F} = \texttt{MLP}([&\mathbf{F}^{A}\oplus\mathbf{F}^{B}\oplus\mathbf{X}^A_{\texttt{specific}}\\
    &\oplus\mathbf{X}^{A}_{\texttt{shared}}\oplus\mathbf{X}^B_{\texttt{specific}}\oplus\mathbf{X}^{B}_{\texttt{shared}}]), \\
    \end{aligned}
    \label{eq:12}
    \end{equation}

    where the $\texttt{MLP}(\cdot)$ means multi-layer feed-forward layer, $\oplus$ denotes the concatenation operator, and the $\mathbf{F}^{A}, \mathbf{F}^{B}, \mathbf{F} \in\mathbb{R}^{T\times D}$ are fused sequence representations.
    Next, we encode the user’s holistic preferences as:
    \begin{equation}
        \mathbf{f} = \frac{1}{T}\sum_{t=0}^{T}\mathbf{F}_{t}, \label{eq: 13}
    \end{equation}
    where the $\mathbf{f} \in\mathbb{R}^{D}$ is the final sequence representation fused domain-specific and domain-shared factor.

\subsection{Prior and Posterior Distributions Alignment}
\begin{sloppypar} 
    Up to now, we have introduced the sequence modeling procedure, in this section, we will dive into NP principle modeling.
    Specifically, in CDSRNP, we formulate each sample in support set $\mathcal{S}$ and query set $\mathcal{Q}$ as: $(X^A,X^B,v_{T+1},y))$, where ($v_{T+1}$ is an arbitrary item belongs to domain $A$ or $B$, $y$ is a 0/1 label to describe the $v_{T+1}$ is a positive or negative item.
    For each sample, we first apply Eq.(\ref{eq:10}) and Eq.(\ref{eq:12}) to generate $\{\mathbf{f}_1, \dots, \mathbf{f}_{N_s}\}$ for support set, and $\{\mathbf{f}_1, \dots, \mathbf{f}_{N_q}\}$ for query set.
    Afterward, we follow the NP principle to encode the encoded user interests, candidate items, and the label of the candidate item.
    For each sample in support set $\mathcal{S}$ and query set $\mathcal{Q}$, we have:
    \begin{equation}
        \mathbf{r}=\texttt{MLP}(\left[\mathbf{f}\oplus\bar{\mathbf{v}}_{T+1}\oplus y\right]),
    \end{equation}
    where $\mathbf{r} \in \mathbb{R}^{D}$ is the sample representation in NP, and the $\bar{\mathbf{v}}_{T+1}$ is candidate item embedding selected from $\bar{\mathbf{V}}$.
    Based on the sample representation, we could aggregate the prior support distribution and the posterior query distribution:
    To satisfy the exchangeability of the neural process \cite{oksendal2013stochastic}, in CDSRNP, we also utilize a straightforward \texttt{Mean} function to obtain the support set representation and query set representation.
    \begin{equation}
        \begin{aligned}
        \mathbf{r}_s = &\texttt{Mean}(\{\mathbf{r}_1, \dots, \mathbf{r}_{N_s}\}),\\
        \mathbf{r}_q = &\texttt{Mean}(\{\mathbf{r}_1, \dots, \mathbf{r}_{N_q}\}),\\
        \end{aligned}
    \end{equation}
    where the $\mathbf{r}_s$ and $\mathbf{r}_q$ describes the support/query representation, respectively.
    To better sample from the stochastic process and to update the parameters using a neural network, we utilize the VAE's reparameterization technique \cite{kingma2013auto} to project the $\mathbf{r}_s$ and $\mathbf{r}_q$ as a normal distribution variable $\mathbf{z}$.
    Taking the support set representation $\mathbf{r}_s$ as example, we have:
\begin{equation}
    \begin{aligned}
        \hat{\mathbf{r}}_{s} = \texttt{ReLU}(\mathbf{W}_r&\mathbf{r}_s), \quad 
        \bm{\mu}_{s} = \mathbf{W}_{\mu}\hat{\mathbf{r}}_{s}, \quad \log\bm{\sigma}_{s} = \mathbf{W}_{\sigma}\hat{\mathbf{r}}_{s},\\
        &\mathbf{z}_s=\bm{\mu}_s+\bm{\epsilon}\odot\bm{\sigma}_s,\quad{\bm{\epsilon}}\sim\mathcal{N}(0,\mathbf{I}),\\
    \end{aligned}
    \label{setlant}
\end{equation}
    where variable $\mathbf{z}_s \in\mathbb{R}^D$ is used to express support set distribution which recorded a sampled users' behaviors pattern, $\mathbf{W}_r, \mathbf{W}_\mu, \mathbf{W}_\sigma\in\mathbb{R}^{D\times D}$ are learnable parameters,  $\odot$ denotes the element-wise product operation, $\sigma$ is a sampled random noise from standard normal distribution. 
    Similarly, the Eq.(\ref{setlant}) also holds to query set to generate query set distribution variable $\mathbf{z}_q$.

    According to the support set variable $\mathbf{z}_s$ and query set variable $\mathbf{z}_q$, we could align them to maximize the users' behaviors distribution across domains:
    \begin{equation}
        \mathcal{L}_{KL}=\mathrm{KL}(q(\mathbf{z}_q|{\mathcal{Q}})||q(\mathbf{z}_s|{\mathcal{S}})).
        \label{eq: KL_loss}
    \end{equation}
    Actually, NP imposes $\mathcal{L}_{KL}$ for the consistency condition in the approximated stochastic process, which practically minimizes the KL divergence between the approximate posterior of the query set $\mathcal{Q}$ and conditional prior of the support set $\mathcal{S}$.
    It is worth noting that our support set is overlapped users, which involves items in different domains at the same time.
    Hence, the non-overlapped user's items in the query set could also fully interact with another item, which could contribute to cross-domain knowledge mining.
    
    \subsection{Interest Adaptive Layer}
    On top of the support/query set variable $\mathbf{z}_s$ and $\mathbf{z}_q$, next we could utilize them as auxiliary general cross-domain information to help our query set sample prediction.
    However, considering the support set size $N_s$ is always a small number, thus the latent variable $\mathbf{z}_s$/$\mathbf{z}_q$ is highly abstract the support set, may provide unstable and coarse-grained cross-domain information.
    Therefore, we further design a fine-grained interest adaptive layer to identify the personalized highly related support users’ interests to transfer beneficial information for enhancing query user prediction
    Specifically, to provide more detailed support set information for each user in the query set, we conduct the cross-attention on sequence representation $\mathbf{f}$ to capture the fine-grained adaptive information from the support set. 
    \begin{equation}
    \begin{aligned}
        \mathbf{f}_s = \texttt{Attention}(\mathbf{f}\mathbf{W}^Q,
        \{\mathbf{f}_{1},\dots, \mathbf{f}_{N_s}\}\mathbf{W}^K,
        \{\mathbf{f}_{1},\dots, \mathbf{f}_{N_s}\}\mathbf{W}^V ),\\
    \end{aligned}
    \end{equation}
    where $\mathbf{f}$ is a arbitrary sequence representation in query set, the $\{\mathbf{f}_{1},\dots, \mathbf{f}_{N_s}\}$ are support set sequence representation, $\mathbf{W}^{Q},\mathbf{W}^{K},\mathbf{W}^{V}\in\mathbb{R}^{D\times D}$
    are parameters to be learned and Attention function.
    Furthermore, note that the fine-grained cross-domain mechanism also provides an opportunity for the non-overlapped user's sequence in the query set to connect other domain item sequences in support, which is beneficial to contribute to cross-domain knowledge mining.

\subsection{Training and Inference}
    For prediction, we combine the support/query user's interest distribution variable $\mathbf{z}_s$/$\mathbf{z}_q$, and detailed variables $\mathbf{f}_s$ obtained through cross-attention for each query set user.
    This combined representation is then passed through linear layers to produce the feature vector $d$:
    \begin{equation}
    \begin{aligned}
        \mathbf{d} = &\texttt{MLP}([\mathbf{f}\oplus \mathbf{f_s}\oplus \mathbf{z_q}]) \quad \texttt{if model training},  \\
        \mathbf{d} = &\texttt{MLP}([\mathbf{f}\oplus \mathbf{f_s}\oplus \mathbf{z_s}]) \quad \texttt{else model inference},  \\
    \end{aligned}
    \end{equation}
    where the $\mathbf{d}$ is the enhanced user interests representation.
    Note that we input $\mathbf{z_q}$ in the training procedure, but $\mathbf{z_s}$ is the inference procedure; this is because the $\mathbf{z_q}$ is not Observable in the inference procedure.
    Finally, the feature vector $\mathbf{d}$ is concatenated with the predicted item vector $v_{T+1}$ and passed through linear layers to generate the final score for this item, which represents the likelihood of the item.
    \begin{equation}
         \hat{y} = \texttt{Sigmoid}(\texttt{MLP}([\mathbf{d}\oplus\bar{\mathbf{v}}_{T+1}])),
    \end{equation}
    where the $\hat{y}\in (0,1)$ is the prediction score of a data sample $(X^A,X^B,v_{T+1},y))$.
\end{sloppypar}

\subsection{Model Optimization}
    To train our model, the loss function consists of three terms: the desired log-likelihood to achieve CDSR, the KL divergence to impose regularization and regularization terms for model parameters.
    Particularly, we rephrase the likelihood $\mathcal{L}_{rec}$ in Eq.\eqref{eq:5} as a regression-based loss function by MSE loss:
    \begin{equation}
        \begin{aligned}
        \mathcal{L}_{rec}=\frac{1}{|{\mathcal{Q}|}}\sum_{i=1}^{|{\mathcal{Q}}|}(y_{i}-\hat{y}_{i})^{2}.
        \label{eq: rec_loss}
        \end{aligned}
    \end{equation}
    In summary, the overall loss function is defined as follows:
    \begin{equation}
        \mathcal{L}=\mathcal{L}_{rec}+\mathcal{L}_{KL}+
        \lambda\|\Theta\|_2,
        \label{eq:loss}
    \end{equation}
    where $\Theta$ is the set of learnable parameters in model and $\lambda$ as the hyper-parameters for learnable parameters.

\section{EXPERIMENTS}
\subsection{Datasets}
    We selected two cross-domain scenarios from the Amazon dataset \cite{amazon1,amazon2}~\footnote{\url{https://cseweb.ucsd.edu/~jmcauley/datasets/amazon_v2/}}, which are named “Cloth-Sport" and “Phone-Elec".
    We count the data statistics for the two cross-domain scenarios in Table \ref{tab:datasetAmazon}.

\begin{table}[htbp]
    \caption{Statistics on the Amazon datasets.} 
    \label{tab:datasetAmazon}
    \resizebox{\linewidth}{!}{
  \begin{tabular}{|c|c|cc|ccc|c|}
    \hline
    \multicolumn{2}{|c|}{\textbf{Dataset}} &
    \textbf{Users} &
    \textbf{Items} &
    \textbf{Ratings}  & 
    \textbf{\#Overlap}  &
    \textbf{Avg.length}  &
    \textbf{Density}  \\
    \hline 
    \multirow{4}{*}{Amazon} &
    cloth & 27,519 & 9,481 & 161,010 &
    \multirow{2}{*}{16,337} & 4.39 & 0.06\% \\
    & sport & 107,984 & 40,460 & 851,553 & &
    7.58 & 0.02\% \\
    \cline{2-8}
    \multirow{4}{*}{} &
    Phone & 41,829 & 17,943 & 194,121 &
    \multirow{2}{*}{7,857} & 4.53 & 0.03\% \\
    & Elec & 27,328 & 12,655 & 170,426 & &
    6.19 & 0.05\% \\
    \hline
  \end{tabular}
}
\end{table}

\begin{table*}[htbp]
\caption{Experimential results(\%) on the bi-directional Cloth-Sport CDSR scenario with different $\mathcal{K}_u$.}
\label{tab:results1}
\centering
\resizebox{\textwidth}{!}{%
  \begin{tabular}{@{}ccccccccc}
    \toprule
    \multirow{3}{*}{\textbf{Methods}}& 
    \multicolumn{4}{c}{\textbf{Cloth-Domain Recommendation}} &
    \multicolumn{4}{c}{\textbf{Sport-Domain Recommendation}} \\
    \cmidrule(lr){2-9}  &

    \multicolumn{2}{c}{$\mathcal{K}_u=25\%$}& 
    \multicolumn{2}{c}{$\mathcal{K}_u=75\%$}&
    \multicolumn{2}{c}{$\mathcal{K}_u=25\%$}& 
    \multicolumn{2}{c}{$\mathcal{K}_u=75\%$}\\ 
    \cmidrule(lr){2-3}
    \cmidrule(lr){4-5}
    \cmidrule(lr){6-7}
    \cmidrule(lr){8-9} &

    NDCG@10 & HR@10 & NDCG@10 & HR@10 & NDCG@10 & HR@10 & NDCG@10 & HR@10  \\ \midrule
    BERT4Rec \cite{bert4rec} 
    & 1.42$\pm$0.13 & 2.96$\pm$0.15 & 2.28$\pm$0.19 & 4.43$\pm$0.46 
    & 2.96$\pm$0.25 & 5.60$\pm$0.34 & 4.19$\pm$0.18 & 7.77$\pm$0.24 \\ 
    GRU4Rec \cite{hidasi2015session} 
    & 2.03$\pm$0.23 & 4.19$\pm$0.35 & 3.23$\pm$0.15 & 6.18$\pm$0.37 
    & 3.63$\pm$0.22 & 7.35$\pm$0.47 & 4.85$\pm$0.16 & 9.43$\pm$0.20  \\ 
    SASRec \cite{kang2018self} 
    & 2.00$\pm$0.13 & 4.26$\pm$0.23 & 3.34$\pm$0.18 & 6.66$\pm$0.38 
    & 3.69$\pm$0.32 & 7.46$\pm$0.55 & 4.96$\pm$0.30 & 9.69$\pm$0.38 \\ 
    \cmidrule(lr){1-9}

    STAR \cite{sheng2021one} 
    & 1.97$\pm$0.27 & 4.19$\pm$0.39 & 3.37$\pm$0.15 & 6.54$\pm$0.28 
    & 3.40$\pm$0.29 & 7.10$\pm$0.51 & 4.77$\pm$0.18 & 9.29$\pm$0.15 \\ 
    MAMDR \cite{luo2023mamdr} 
    & 2.11$\pm$0.09 & 4.25$\pm$0.17 & 3.44$\pm$0.15 & 6.60$\pm$0.15 
    & 3.53$\pm$0.44 & 7.21$\pm$0.24 & 4.84$\pm$0.14 & 9.33$\pm$0.42 \\ 
    SSCDR \cite{kang2019semi} 
    & 2.02$\pm$0.18 & 4.21$\pm$0.36 & 3.42$\pm$0.24 & 6.57$\pm$0.17 
    & 3.45$\pm$0.24 & 7.14$\pm$0.40 & 4.81$\pm$0.29 & 9.31$\pm$0.57 \\ 
    \cmidrule(lr){1-9}
    
    Pi-Net \cite{ma2019pi} 
    & 1.84$\pm$0.26 & 3.82$\pm$0.38 & 2.77$\pm$0.20 & 5.56$\pm$0.32 
    & 3.37$\pm$0.17 & 7.05$\pm$0.22 & 4.56$\pm$0.25 & 8.81$\pm$0.34 \\ 
    DASL \cite{li2021dual} 
    & 2.21$\pm$0.29 & 4.56$\pm$0.43 & 3.32$\pm$0.17 & 6.62$\pm$0.29 
    & 3.96$\pm$0.16 & 8.11$\pm$0.37 & 4.84$\pm$0.39 & 9.71$\pm$0.51 \\
    C$^{2}$DSR \cite{cao2022contrastive} 
    & 2.27$\pm$0.18 & 4.73$\pm$0.35 & 3.31$\pm$0.07 & 6.62$\pm$0.24 
    & 3.74$\pm$0.26 & 7.99$\pm$0.42 & 5.18$\pm$0.10 & 10.31$\pm$0.07 \\
    \cmidrule(lr){1-9}
    
    BERT4Rec \cite{bert4rec} + AMID \cite{rethinking}
    & 2.99$\pm$0.01 & 5.70$\pm$0.13 & 3.79$\pm$0.15 & 7.09$\pm$0.32 
    & 4.73$\pm$0.29 & 9.30$\pm$0.41 & 5.70$\pm$0.12 & 10.63$\pm$0.36  \\ 
    
    GRU4Rec \cite{hidasi2015session} + AMID \cite{rethinking}
    & 3.10$\pm$0.17 & 5.95$\pm$0.16 
    & 3.94$\pm$0.15 & 7.30$\pm$0.30
    & 4.89$\pm$0.10 & 9.42$\pm$0.28 
    & 5.90$\pm$0.17 & 11.01$\pm$0.17 \\  
    SASRec \cite{kang2018self} + AMID \cite{rethinking}
    & \underline{3.20$\pm$0.22} & \underline{6.14$\pm$0.33}
    & \underline{4.19$\pm$0.10} & \underline{7.62$\pm$0.20} 
    & \underline{4.97$\pm$0.15} & \underline{9.48$\pm$0.22} 
    & \underline{5.96$\pm$0.14} & \underline{11.04$\pm$0.26} \\ 
    \cmidrule(lr){1-9}

    \textbf{CDSRNP(ours)}   
    & \textbf{3.60$\pm$0.21} & \textbf{6.66$\pm$0.48} 
    & \textbf{4.55$\pm$0.23} & \textbf{8.15$\pm$0.37} 
    & \textbf{5.54$\pm$0.19} & \textbf{10.08$\pm$0.25} 
    & \textbf{6.76$\pm$0.27} & \textbf{12.22$\pm$0.13} \\ 
    
    Improvement(\%)
    & 12.50 & 8.46 & 8.59 & 6.95
    & 11.46 & 6.33 & 16.94 & 10.68 \\ 
    \bottomrule
\end{tabular}
}
\footnotesize
\end{table*}

\begin{table*}[ht]
\caption{Experimential results(\%) on the bi-directional Phone-Elec CDSR scenario with different $\mathcal{K}_u$.}
\label{tab:results2}
\centering
\resizebox{\textwidth}{!}{%
  \begin{tabular}{@{}ccccccccc}
    \toprule
    \multirow{3}{*}{\textbf{Methods}} &
    \multicolumn{4}{c}{\textbf{Phone-Domain Recommendation}} &
    \multicolumn{4}{c}{\textbf{Elec-Domain Recommendation}} \\
    \cmidrule(lr){2-9}  &

    \multicolumn{2}{c}{$\mathcal{K}_u=25\%$}& 
    \multicolumn{2}{c}{$\mathcal{K}_u=75\%$}&
    \multicolumn{2}{c}{$\mathcal{K}_u=25\%$}& 
    \multicolumn{2}{c}{$\mathcal{K}_u=75\%$}\\ 
    \cmidrule(lr){2-3}
    \cmidrule(lr){4-5}
    \cmidrule(lr){6-7}
    \cmidrule(lr){8-9} &

    NDCG@10 & HR@10 & NDCG@10 & HR@10 & NDCG@10 & HR@10 & NDCG@10 & HR@10 \\ \midrule
    BERT4Rec \cite{bert4rec} 
    & 6.13$\pm$0.16 & 10.87$\pm$0.24 & 6.20$\pm$0.17 & 11.07$\pm$0.47
    & 8.24$\pm$0.22 & 13.05$\pm$0.33 & 10.58$\pm$0.07 & 17.28$\pm$0.05 \\ 
    GRU4Rec \cite{hidasi2015session} 
    & 6.99$\pm$0.23 & 12.98$\pm$0.39 & 6.98$\pm$0.18 & 12.90$\pm$0.38 
    & 10.11$\pm$0.12 & 16.77$\pm$0.23 & 11.41$\pm$0.08 & 19.00$\pm$0.12  \\ 
    SASRec \cite{kang2018self} 
    & 7.19$\pm$0.19 & 13.32$\pm$0.30 & 7.16$\pm$0.26 & 13.34$\pm$0.34 
    & 1056$\pm$0.18 & 17.77$\pm$0.21 & 11.64$\pm$0.04 & 19.67$\pm$0.18 \\ 
    \cmidrule(lr){1-9}

    STAR \cite{sheng2021one} 
    & 6.96$\pm$0.24 & 13.30$\pm$0.45 & 7.13$\pm$0.18 & 13.71$\pm$0.51
    & 9.72$\pm$0.17 & 16.38$\pm$0.26 & 11.18$\pm$0.13 & 19.00$\pm$0.29 \\ 
    MAMDR \cite{luo2023mamdr} 
    & 7.01$\pm$0.22 & 13.38$\pm$0.42 & 7.19$\pm$0.19 & 13.81$\pm$0.20 
    & 9.79$\pm$0.31 & 16.47$\pm$0.39 & 11.24$\pm$0.47 & 19.08$\pm$0.25 \\ 
    SSCDR \cite{kang2019semi} 
    & 6.99$\pm$0.39 & 13.35$\pm$0.45 & 7.21$\pm$0.35 & 13.83$\pm$0.30 
    & 9.75$\pm$0.27 & 16.45$\pm$0.27 & 11.27$\pm$0.31 & 19.13$\pm$0.33 \\ 
    \cmidrule(lr){1-9}
    
    Pi-Net \cite{ma2019pi} 
    & 7.02$\pm$0.26 & 12.68$\pm$0.50 & 7.11$\pm$0.26 & 12.95$\pm$0.36 
    & 10.00$\pm$0.11 & 16.33$\pm$0.28 & 11.61$\pm$0.06 & 19.34$\pm$0.26 \\ 
    DASL \cite{li2021dual} 
    & 7.10$\pm$0.15 & 13.29$\pm$0.17 & 7.22$\pm$0.13 & 13.22$\pm$0.17 
    & 10.71$\pm$0.13 & 17.98$\pm$0.10 & 11.74$\pm$0.03 & 20.04$\pm$0.21 \\
    C$^{2}$DSR \cite{cao2022contrastive} 
    & 7.54$\pm$0.15 & 14.04$\pm$0.23 & 7.30$\pm$0.19 & 13.96$\pm$0.44 
    & 10.71$\pm$0.13 & 17.98$\pm$0.10 & 11.74$\pm$0.03 & 20.04$\pm$0.21 \\
    \cmidrule(lr){1-9}
    
    BERT4Rec \cite{bert4rec} + AMID \cite{rethinking}
    & 6.83$\pm$0.13 & 11.81$\pm$0.25 & 6.84$\pm$0.14 & 12.11$\pm$0.34 
    & 8.41$\pm$0.13 & 13.38$\pm$0.16 & 11.23$\pm$0.17 & 18.39$\pm$0.34 \\  
    GRU4Rec \cite{hidasi2015session} + AMID \cite{rethinking}
    & \underline{8.33$\pm$0.09} & \underline{15.49$\pm$0.50} 
    & \underline{8.34$\pm$0.21} & \underline{15.29$\pm$0.29} 
    & \textbf{11.74$\pm$0.07} & \underline{19.50$\pm$0.08} 
    & \underline{12.53$\pm$0.11}  & \underline{20.85$\pm$0.20} \\  
    SASRec \cite{kang2018self} + AMID \cite{rethinking}
    & 8.20$\pm$0.10 & 14.99$\pm$0.16 & 8.32$\pm$0.20 & 14.96$\pm$0.15 
    & 11.71$\pm$0.23 & 19.28$\pm$0.23 & 12.52$\pm$0.13 & 20.79$\pm$0.20 \\ 
    \cmidrule(lr){1-9}

    \textbf{CDSRNP(ours)}   
    & \textbf{8.87$\pm$0.37} & \textbf{16.07$\pm$0.56} 
    & \textbf{8.93$\pm$0.22} & \textbf{16.37$\pm$0.51}   
    & \underline{11.36$\pm$0.19} & \textbf{19.51$\pm$0.27} 
    & \textbf{12.56$\pm$0.14} & \textbf{21.30$\pm$0.32} \\ 
    
    Improvement(\%)
    & 6.48 & 3.74 & 7.07 &  7.06
    & -3.23 & 0.05 & 0.24 & 2.16 \\ 
    \bottomrule
\end{tabular}
}
\footnotesize
\end{table*}

\subsection{Experimental Setting}
    Our experiments followed the previous experimental setup \cite{cao2022contrastive,xu2023neural}.
    We filtered the dataset for users with less than 10 interactions and products with less than 5 interactions.
    In each cross-domain scenario, we allocate 80\% of the dataset for training, 10\% for validation, and 10\% for testing.
    We introduce the non-overlapping ratio $\mathcal{K}_u$ \cite{rethinking}, which controls the percentage of overlapping users in the dataset, to simulate different scenarios.
    We set $\mathcal{K}_u$ to $25\%$ and $75\%$ to simulate scenarios.
    The larger the $\mathcal{K}_u$, the higher the share of non-overlapping users.
    We control the training and validation sets via $\mathcal{K}_u$, but use all users in the test set.

    \subsubsection{Evaluation Metrics}
    We evaluate our model using a standard approach\cite{krichene2020sampled,zhou2018deep}: combining 1 positive item with 999 randomly selected negative items for ranking tests. Performance is measured using $NDCG@10$ and $HR@10$ metrics, common in CDSR research. 
    All experiments are repeated five times, with results reported as means and variances for statistical robustness.

    \subsubsection{Compared Methods} 
    \begin{sloppypar} 
    We evaluated our approach against three categories of recommendation systems:
    Single Domain Sequential Recommendation Methods:
    BERT4Rec\cite{bert4rec}, GRU4Rec\cite{hidasi2015session}, SASRec\cite{kang2018self}
    Conventional Cross-Domain Recommendation Methods:STAR\cite{sheng2021one}, MAMDR\cite{luo2023mamdr}, SSCDR\cite{kang2019semi}
    Cross-Domain Sequential Recommendation Methods:
    Pi-Net\cite{ma2019pi}, DASL\cite{li2021dual},
    C$^2$DSR\cite{cao2022contrastive},
    AMID\cite{rethinking}
    \end{sloppypar}

    \subsubsection{Implementation and Hyperparameter Setting}
    To ensure a fair comparison between different approaches, we standardized the hyperparameters across all methods. 
    We set the embedding dimension to 128 and used a negative sampling ratio of 1 for training, 199 for validating and 999 for testing, with negative samples in the query set and only positive samples in the support set. 
    The Adam optimizer was employed to update all parameters.
    For baseline comparisons, we adopted the hyperparameter values reported in the original literature.
    In our proposed model, we fixed $\lambda$ at 1e-5 to regularize model parameters. 
    The learning rate was set to 1e-5 for the “Phone-Elec” dataset and 2e-5 for the “Cloth-Sport” dataset.
    We trained the model for 50 epochs, varying the interaction sequence length from $[15,25,35,45]$. The support set size ranged from $[10,20,30,40]$, while the query set size was fixed at 20 (comprising 10 positive and 10 negative samples). 

\subsection{Performance Comparisons on CDSR}
    We count our experimental performance with the comparison models under two cross-domain scenarios and different proportions of non-overlapping users in Table \ref{tab:results1} and Table \ref{tab:results2}.
    The bolded bold is the best result, and the lower crossover is the second best result.
    Our model is the best in most experimental results, and some of the experimental results $NDCG@10$ are improved by 16.94\% compared with the second place in the domain.
    We can observe that in most of the experimental results, there is more enhancement in the scenario with $\mathcal{K}_u=75\%$ compared to domain $\mathcal{K}_u=25\%$.
    we can observe that in scenarios where the proportion of non-overlapping users is higher, it is important to go for mining the information of non-overlapping users to allow the model to learn better.
    The experimental results show that our method CDSRNP can unleash the potential of non-overlapping users, no longer relying solely on overlapping users, removing the bias and allowing the model to learn knowledge better.

\subsection{Ablation Study}
    \subsubsection{Various Model Variants}
    \begin{sloppypar}
    To further assess our model's ability to leverage non-overlapping users, we analyzed their performance in two domains with higher proportions of such users. 
    Table \ref{tab:single domain user} demonstrates that our method significantly outperforms the strongest baseline across all metrics, indicating that CDSRNP more effectively utilizes non-overlapping users and enhances their recommendations.
    We conducted an ablation study to evaluate the effectiveness of key components in our approach. 
    We tested variants using (1) a single embedding and sequence encoder and (2) removing the interest adaptive layer. 
    (3) We also examined the impact of user identity in the support set by random sampling among all users. 
    Table \ref{tab:variants} presents the comparative results under these different configurations.
    Our full CDSRNP model consistently outperforms its variants, highlighting the positive contributions of each module. 
    Moreover, the use of overlapping users in the support set is beneficial to facilitate better knowledge transfer and unleash the potential of non-overlapping users.
    
\begin{table}[htbp]
    \caption{Experimental results for single domain users}
    \label{tab:single domain user}
    \resizebox{\linewidth}{!}{
    \begin{tabular}{cccccc}
        \hline
        \multirow{2}{*}{\textbf{Method}} &
        \multirow{2}{*}{$\mathcal{K}_u$} &
        \multicolumn{2}{c}{\textbf{Sport-Domain}} &
        \multicolumn{2}{c}{\textbf{Elec-Domain}} \\
        \cmidrule(lr){3-4} \cmidrule(lr){5-6}
         &  & NDCG@10 & HR@10 & NDCG@10 & HR@10 \\
        \hline

        \multirow{2}{*}{SASRec + AMID} 
        & $25\%$ 
        & 3.72$\pm$0.22 & 7.43$\pm$0.33 
        & 9.77$\pm$0.06 & 16.39$\pm$0.14  \\ 
        & $75\%$
        & 5.56$\pm$0.17 & 10.70$\pm$0.23 
        & 11.16$\pm$0.17 & 18.89$\pm$0.20  \\
        \hline
                
        \multirow{2}{*}{CDSRNP} 
        & $25\%$ 
        & 4.46$\pm$0.18 & 8.43$\pm$0.34
        & 10.31$\pm$0.14 & 17.97$\pm$0.25  \\ 
        & $75\%$
        & 6.71$\pm$0.31 & 12.20$\pm$0.53 
        & 11.98$\pm$0.04 & 20.39$\pm$0.25  \\
        \hline
    \end{tabular}
    }
\end{table}

\begin{table}[b!]
    \caption{Experimental results for various model variants}
    \label{tab:variants}
    \resizebox{\linewidth}{!}{
    \begin{tabular}{cccccc}
        \hline
        \multirow{2}{*}{\textbf{Method}} &
        \multirow{2}{*}{$\mathcal{K}_u$} &
        \multicolumn{2}{c}{\textbf{Phone-Domain}} &
        \multicolumn{2}{c}{\textbf{Elec-Domain}} \\
        \cmidrule(lr){3-4} \cmidrule(lr){5-6}
         &  & NDCG@10 & HR@10 & NDCG@10 & HR@10 \\
        \hline
        \multirow{2}{*}{one embedding} 
        & $25\%$ 
        & 8.30$\pm$0.27 & 15.32$\pm$0.83
        & 9.94$\pm$0.16 & 16.99$\pm$0.24  \\ 
        & $75\%$
        & 8.43$\pm$0.19 & 15.45$\pm$0.65 
        & 11.96$\pm$0.83 & 20.17$\pm$0.30  \\
        \hline
        
        \multirow{2}{*}{no adaptive layer} 
        & $25\%$ 
        & 8.61$\pm$0.26 & 15.45$\pm$0.29 & 11.17$\pm$0.21 & 19.42$\pm$0.31  \\ 
        & $75\%$
        & 8.68$\pm$0.16 & 15.88$\pm$0.53 & 11.88$\pm$0.59 & 20.25$\pm$0.88  \\
        \hline
        
        \multirow{2}{*}{all user} 
        & $25\%$ 
        & 8.58$\pm$0.38 & 15.53$\pm$0.59 
        & 11.24$\pm$0.12 & 19.38$\pm$0.17  \\ 
        & $75\%$
        & 8.66$\pm$0.30 & 15.86$\pm$0.30 
        & 12.45$\pm$0.10 & 21.18$\pm$0.20  \\
        \hline
                
        \multirow{2}{*}{overlap user} 
        & $25\%$ 
        & 8.87$\pm$0.37 & 16.07$\pm$0.56
        & 11.36$\pm$0.19 & 19.51$\pm$0.27  \\ 
        & $75\%$
        & 8.93$\pm$0.22 & 16.37$\pm$0.51 
        & 12.56$\pm$0.14 & 21.30$\pm$0.32  \\
        \hline
    \end{tabular}
    }
\end{table}

\subsubsection{Impact of Hyper-parameters}
\begin{figure}
	\centering
	\begin{minipage}{0.49\linewidth}
		\centering
		\includegraphics[width=0.9\linewidth]{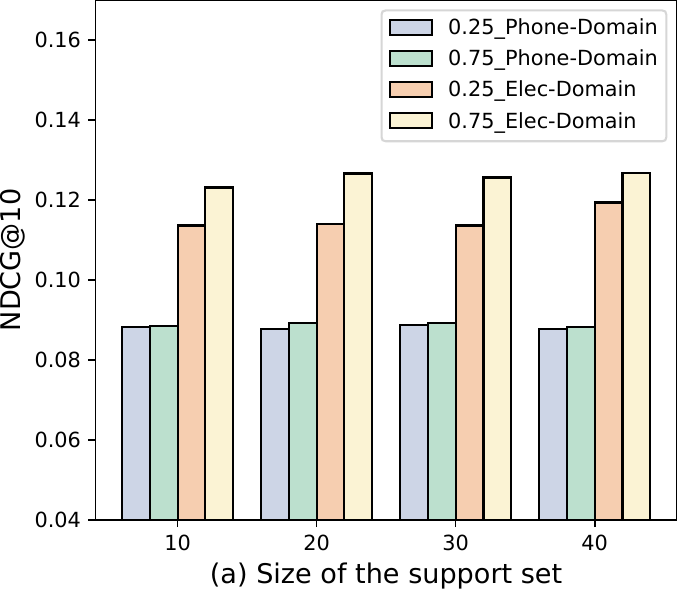}
	\end{minipage}
	\begin{minipage}{0.49\linewidth}
		\centering
		\includegraphics[width=0.9\linewidth]{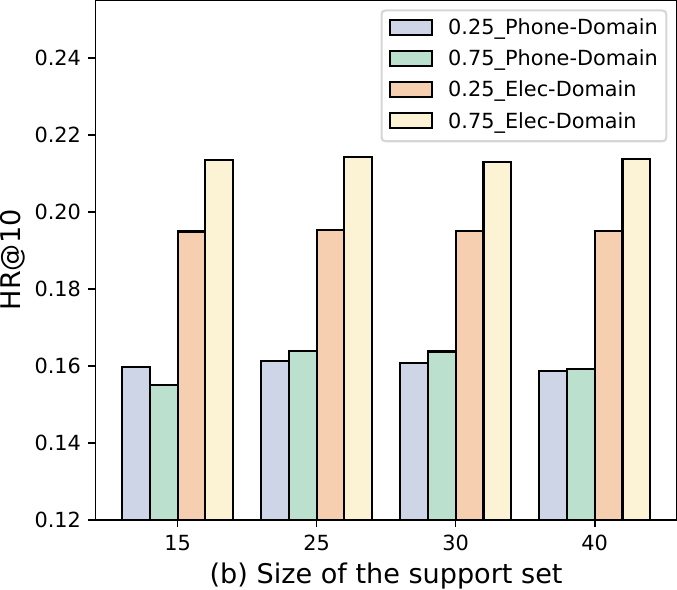}
	\end{minipage}
	\begin{minipage}{0.49\linewidth}
		\centering
		\includegraphics[width=0.9\linewidth]{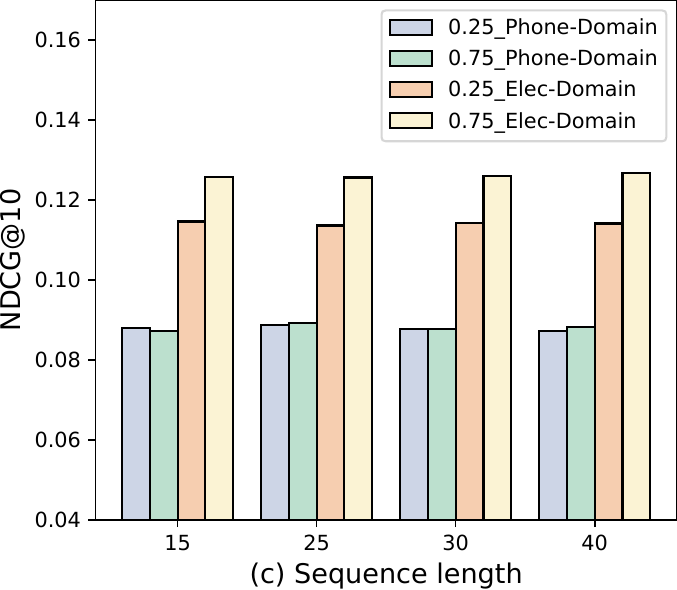}
	\end{minipage}
	\begin{minipage}{0.49\linewidth}
		\centering
		\includegraphics[width=0.9\linewidth]{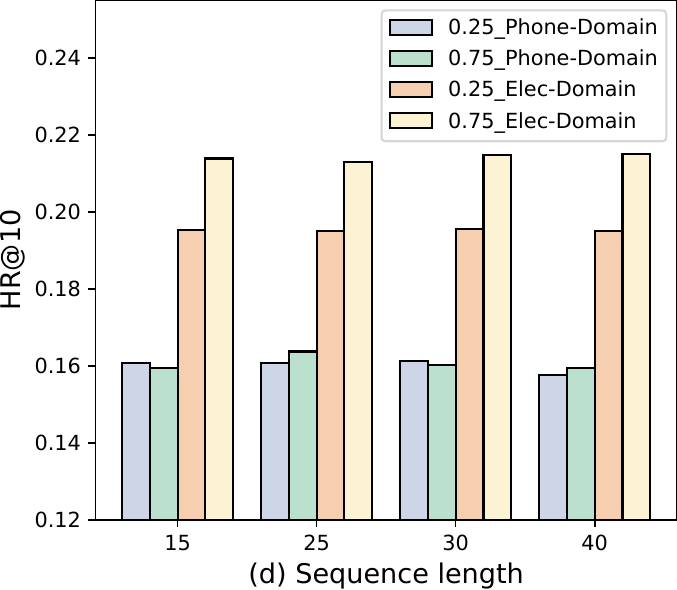}
	\end{minipage}
    \caption{Performance of different hyper-parameters}
    \label{hyper}
    \Description{}
\end{figure}
    We conducted a series of experiments to explore various parameters in the model and their impact on performance in Fig. \ref{hyper}.
    (1) Size of the support set:
    we investigated how different support set sizes affect the model's performance. 
    (2) Sequence length:
    we examined the impact of varying the maximum length of user interactions on the model's performance.
    Interestingly, our findings show that even with relatively short interaction lengths, the model can still provide reasonably good recommendations. 
    Furthermore, changes in the size of the support set did not significantly alter the model's recommendation effectiveness, indicating that the model is robust to variations in these hyperparameters.

\end{sloppypar}

\section{Related Work}
    \textbf{Single-Domain Sequential Recommendation} This field has seen significant advancements with models like BERT4Rec \cite{bert4rec}, which utilizes bidirectional self-attention for item prediction. 
    GRU4Rec \cite{hidasi2015session} adapted RNNs for sparse sequential data, introducing novel ranking loss functions. 
    SASRec \cite{kang2018self} is based on the sequential model, using self-attention modeling user interaction sequences to predict the next item.\\
    \textbf{Conventional Cross-Domain Recommendation} STAR \cite{sheng2021one}: Trains a single model for multiple domains using shared and domain-specific factorized networks.
    MaMDR \cite{luo2023mamdr}: Presents a model-agnostic framework for multi-domain recommendation, addressing data distribution variations and inter-domain conflicts.
    SSCDR \cite{kang2019semi}: Focuses on cold-start users, using a semi-supervised mapping approach to learn cross-domain relationships with limited labeled data.
    There are also some studies \cite{unicdr, disencdr, cdrib, cao2024unified}\\
    \textbf{Cross-Domain Sequential Recommendation} Pi-Net \cite{ma2019pi}: Generates recommendations for two domains simultaneously, sharing user behaviors at each timestamp to improve accuracy in cross-domain scenarios.
    DASL \cite{li2021dual}: Considers bidirectional latent relations of user preferences across domain pairs, featuring dual learning mechanisms for cross-domain CTR predictions.
    C2DSR \cite{cao2022contrastive}: Jointly learns single- and cross-domain user preferences by leveraging intra- and inter-sequence item relationships.
    AMID \cite{rethinking}: An adaptive multi-interest framework and introduces a cross-domain debiasing estimator to reduce the estimation bias.
    % There are also some studies \cite{wang2023linsatnet,wang2020combinatorial,scarselli2008graph,yang2023debiased}\\
    \textbf{Neural Processes (NP)} NP offers an efficient approach to modeling function distributions using neural network architectures.
    Initially applied to image-related tasks \cite{garnelo2018conditional,garnelo2018neural,wang2022np}, NPs have been adapted for recommendation systems. 
    Enhanced variants such as ANP \cite{kim2019attentive} and BNP \cite{lee2020bootstrapping} have improved reconstruction quality and stochasticity modeling. 
    Recent applications in recommendation systems, exemplified by TANP \cite{lin2021task} and CDRNP \cite{li2024cdrnp}, address cold-start problems but primarily focus on single-domain contexts.
    In addition to this, there are a number of studies \cite{finn2017model,munkhdalai2017meta}

\section{CONCLUSION}
    This paper introduces CDSRNP, a novel framework for CDSR that addresses the challenge of leveraging non-overlapping user behaviors to enhance recommendation effectiveness. 
    Traditional CDSR methods struggle with non-overlapping users, as the absence of cross-domain interactive behaviors hinders the capture of inter-domain item connections.
    Our approach employs neural processes to bridge this gap, enabling the capture of cross-domain knowledge from non-overlapping user interactions.
    CDSRNP implements a meta-learning workflow and a fine-grained interest adaptive layer to enhance user predictions.
    Extensive experiments on real-world datasets demonstrate CDSRNP's significant performance improvements over state-of-the-art baselines, validating our approach and highlighting its potential to address data sparsity in recommendation systems. 
    Future research could explore further refinements and investigate its applicability in diverse recommendation scenarios.

\bibliographystyle{ACM-Reference-Format}
\bibliography{sample-base}

\end{document}